\documentclass[12pt]{article}
\usepackage{amsmath,amsfonts}
\def\ie{{\em i.e.\,}}
\def\bra{\langle}
\def\ket{\rangle}
\begin{document}
\title{Quantum neural network}
\author{M.V.Altaisky \\
\small Joint Institute for Nuclear 
Research, Dubna, 141980, RUSSIA;  
 \\ \small Space Research Institute RAS, Profsoyuznaya 84/32, Moscow, 
117997, RUSSIA}
\date{}
\maketitle
\begin{abstract}
It is suggested that a quantum neural network (QNN), a type of 
artificial neural network,  can be built 
using the principles of quantum information processing. The input and 
output qubits 
in the QNN can be implemented  by optical modes with different polarization, 
the weights of the QNN can be implemented by optical beam splitters and 
phase shifters.
\end{abstract}
Since it was first proposed by Feynman \cite{Feynman1982}, that quantum 
mechanics might be more powerful computationally than a classical Turing 
machine, we have heard a lot of quantum computational 
networks \cite{Deutch1989}, quantum cellular automata \cite{Feynman1986}, 
but only a little about quantum neural networks \cite{vlasov}. 
The possible reason for the omni-penetrating ideas of quantum information 
processing (QIP) to avoid the field of artificial neural networks (ANN), 
is the 
presence of a nonlinear activation function in any ANN. For very similar 
reason, a  need for nonlinear couplings between optical modes was the main 
obstacle for building a scalable optical QIP system.   

It was shown recently \cite{KLM01}, that quantum 
computation on optical modes using only beam splitters, phase shifters, 
photon sources and photo detectors is possible. Accepting the ideas of \cite{KLM01}, 
we just assume the existence of a qubit 
\begin{equation}
|x\ket = \alpha |0\ket + \beta |1\ket, 
\label{qubit}
\end{equation}   
where $|\alpha|^2+|\beta|^2=1$, with the states $|0\ket$ and $|1\ket$ are 
understood as  different polarization states of light. 

Let us consider a {\em perceptron}, \ie  the system with $n$ 
input channels $x_1,\ldots,x_n$ and one output channel $y$. The output of 
a classical perceptron \cite{MP1969} is 
\begin{equation}
y = f(\sum_{j=1}^n w_j x_j),
\label{cl-pct}
\end{equation} 
where $f(\cdot)$ is the perceptron activation function and $w_j$ are the 
weights tuning during learning process. 

The perceptron learning algorithm works 
as follows:
\begin{enumerate}
\item The weights $w_j$ are initialized to small random numbers.
\item A {\em pattern} vector ($x_1,\ldots,x_n$) is presented to the 
perceptron and the output $y$ generated according to the rule \eqref{cl-pct}
\item The weights are updated according to the rule
\begin{equation}
w_j(t+1) = w_j(t) + \eta(d-y)x_j,
\label{cl-learn}
\end{equation}
where $t$ is discreet time, $d$ is the desired output provided for teaching  
and $0<\eta<1$ is the step size.
\end{enumerate} 

It  will be hardly possible to construct an exact analog of the nonlinear 
activation function $f$, like sigmoid  and other functions of common use 
in neural networks, but we will show that the leaning rule of the type 
\eqref{cl-learn} is possible for a quantum system too. 

Let us consider a quantum system with $n$ inputs $|x_1\ket,\ldots,|x_n\ket$ 
of the form \eqref{qubit}, and the output $|y\ket$ derived by the rule 
\begin{equation}
|y\ket = \hat F \sum_{j=1}^n \hat w_j |x_j\ket,
\label{q-pct}
\end{equation} 
where $\hat w_j$ become $2\times2$  matrices acting on the basis 
($|0\ket,|1\ket$),  combined 
of phase shifters $e^{\imath\theta}$ and beam splitters, and possibly 
light attenuators, 
(cf. eq.1 from \cite{KLM01}); $\hat F$ is an unknown operator that can 
be implemented by the network of quantum gates. 

Let us consider the simplistic case with $\hat F=1$ being the identity 
operator. The output of the quantum perceptron at the time $t$ 
will be 
\begin{equation}
|y(t)\ket = \sum_{j=1}^n \hat w_j(t) |x_j\ket.
\end{equation}
In analogy with classical case \eqref{cl-learn}, let us provide 
a learning rule 
\begin{equation}
\hat w_j(t+1) = \hat w_j(t) + \eta (|d\ket - |y(t)\ket)\bra x_j |,
\label{q-learn}
\end{equation}
where $|d\ket$ is the desired output.

It is easy to show now, that the learning rule \eqref{q-learn} drives 
the  quantum perceptron into desired state $|d\ket$ used for teachning.
In fact , using the rule \eqref{q-learn} and taking the module-square 
difference of the  real and desired outputs, we yield 
\begin{equation}
\||d\ket - |y(t+1)\ket\|^2 =  
 \||d\ket-\sum_{j=1}^n \hat w_j(t+1)|x_j\ket\|^2 
=(1-n\eta)^2 \||d\ket - |y(t)\ket\ \|^2
\end{equation}  
For small $\eta$ ($0<\eta<1/n$) and normalized input states $\bra x_j|x_j\ket=1$ the 
result of iteration converges to the desired state $|d\ket$. The whole 
network can be then composed from the primitive elements usinng the 
standard rules of ANN architecture.

The learning rule \eqref{q-learn} may cause questions, for it 
does not observe the unitarity in general (in contrust, say, to the ideas 
proposed in \cite{vlasov}), and thus 
not only phase rotations but also light attenuation may take place.  
This a point for future consideration: formally we can observe the 
unitarity by substituting the learning rule \eqref{q-learn} by 
one acting on unit circle basis $e^{\imath m \theta}$, but the model with 
attenuation of certain 
weight in QNN seems more reasonable for learn-ability of the network, and 
more perspective for simmulation on classical computer. 

The idea, this note was inspired by,  was to suggest a neural network 
model, which  takes into account the phase of the signal, rather than only 
the  amplitude, as existing ANN. The idea seems to be very close no  real 
biological neural nets, where the neurons are sensitive to the phase of the 
signal rather than the amplitude alone. In the later case, the matrix weights 
$\hat w_j$ could be understood as complex impedances that attenuate 
the signal and change its phase. 

\centerline{***}
The author is thankful tp Dr. A.Vlasov for useful references and comments.


\newpage
\begin{thebibliography}{9}

\bibitem {Feynman1982} Feynman, R.P. {\em Simulating physics with computers}, 
          Int. J.Theor.Phys. {\bf 21}(1982)467-488.
\bibitem {Feynman1986} Feynman, R.P. {\em Quantum Mechanical Computers}, 
          Found. Phys. {\bf 16}(1986)507-531.
\bibitem {Deutch1989} Deutch, D. {\em Quantum computational networks}, 
          Proc. Roy. Soc. Lond. {\bf A439}(1992)553-558. 
\bibitem {vlasov} Vlasov, A.Yu. 
         {\em Quantum computations and images recognition}, quant-ph/9703010; 
         {\em Analogues quantum computers for data analysis} quant-ph/9802028. 
\bibitem {KLM01}  E.Knill, R.Laflamme and G.J.Milburn. {\em A scheme for 
         efficient quantum computation with linear optics}, 
         Nature {\bf 409}(2001)46-57.
\bibitem {MP1969} M.Minsky and P.Papert, {\em Perceptrons: An Introduction 
         to Computational Geometry}, MIT Press, Cambridge, Mass. 1969. 
\end{thebibliography}
\end{document}